\documentclass[superscriptaddress,twocolumn,nofootinbib]{revtex4}
\usepackage{bm,graphicx}
\usepackage{hyperref}

\begin{document}

\title{Electronic states and persistent currents in nanowire quantum ring}

\author{I.A. Kokurin}
\email[E-mail:]{kokurinia@math.mrsu.ru} \affiliation{Institute of
Physics and Chemistry, Mordovia State University, 430005 Saransk,
Russia} \affiliation{Ioffe Institute, 194021 St. Petersburg, Russia}

\begin{abstract}
The new model of a quantum ring (QR) defined inside a nanowire (NW)
is proposed. The one-particle Hamiltonian for electron in
[111]-oriented NW QR is constructed taking into account both Rashba
and Dresselhaus spin-orbit coupling (SOC). The energy levels as a
function of magnetic field are found using the exact numerical
diagonalization. The persistent currents (both charge and spin) are
calculated. The specificity of SOC and arising anticrossings in
energy spectrum lead to unusual features in persistent current
behavior. The variation of magnetic field or carrier concentration
by means of gate can lead to pure spin persistent current with the
charge current being zero.
\end{abstract}

\date{\today}

\maketitle

\section{Introduction}

The recent progress in nanowire (NW) growth technology, in
particular, the possibility of radial (core-shell) \cite{Funk2013}
and axial \cite{Nylund2016} heterostructure growth, leads to the
opportunity to fabricate various NW-based structures, e.g. so-called
NW-quantum dots (quantum dot inside NW) and other more complex ones.
We suppose that the quantum ring (QR) can be grown in NW by similar
way (see Fig.~\ref{fig1}a), e.g. using well-known hetero-pair ${\rm
GaAs-Al_xGa_{1-x}As}$ with n-doped Al-rich barriers.

Alternatively, the tubular electron gas (TEG) formed close to
InAs-NW surface \cite{Hernandez2010} can be electrostatically
confined to form QR, by analogy with confinement of carriers in
carbon nanotubes to ring geometry~\cite{Bulaev2008}. Both mentioned
structures significantly differ from known QRs defined in
two-dimensional electron gas (2DEG) structures. Especially, the
effects that are due to spin-orbit coupling (SOC) will be different.

QRs constitute the polygon for study such coherent effects as
Aharonov-Bohm (AB) one \cite{Aharonov1959} and persistent currents
(PCs) \cite{Buttiker1983}. Usually zero-temperature PC can be found
by using the well-known equation

\begin{equation}
\label{PC} I=-c\sum_n\frac{\partial E_n}{\partial\Phi},
\end{equation}
where $E_n$ is the energy levels in the ring, $\Phi$ is the magnetic
flux through the ring and the summation is over all occupied states.

The presence of SOC leads to modification of energy spectrum and as
a consequence to PC modification. The spectrum of thin
(one-dimensional) QR defined in 2DEG structure with Rashba SOC
\cite{Bychkov1984} is well known \cite{Chaplik1995,Meijer2002}.
However, the lack of inversion center in host semiconductor material
leads to other type of spin-splitting \cite{Dresselhaus1955} (known
as Dresselhaus SOC), that appear in a new light in low-dimensional
structures \cite{Dyakonov1986} and particularly in QR-structures
\cite{Sheng2006}.

Here we will study only features of PCs that are due to specifics of
spectrum and SOC, and neglect the disorder and interaction effects.
We present the model of thin QR which is confined inside
[111]-oriented NW (it is the usual growth direction for NWs of ${\rm
A_{III}B_V}$ materials with zinc-blende lattice). It should be
noted, that SOC in NW QR sufficiently differs from SOC in planar QR,
where effective Rashba field is constant at each point, and
Dresselhaus SOC is different as well due to another crystallographic
orientation. It is worth noted, that there is no difference between
AB-flux and homogeneous magnetic field for model of thin
one-dimensional QR if $g=0$. However, it seems to be unreal to
realize AB-flux through the NWQR, and we discuss here only the case
of homogeneous magnetic filed.

\section {Model and Hamiltonian}

Using one-particle Hamiltonian of NW with TEG
\cite{Kokurin2014,Kokurin2015}, that takes into account both Rashba
and $k$-linear Dresselhaus SOC, after dimension quantization along
NW axis, we find the following effective-mass Hamiltonian for
carriers in NWQR

\begin{eqnarray}
\label{Hamiltonian} \nonumber
H=\frac{\hbar^2K_\varphi^2}{2m}+\alpha\sigma_zK_\varphi-\frac{\beta}{2\sqrt
3}\left(\sigma_rK_\varphi-\frac{i}{2r_0}\sigma_\varphi\right)\\
+\sqrt\frac32\beta\sigma_z\left(\sin3\varphi
K_\varphi-\frac{3i}{2r_0}\cos
3\varphi\right)+\frac12g\mu_B\sigma_zB,
\end{eqnarray}
where the total Hamiltonian consists of kinetic term, Rashba SOC,
isotropic and anisotropic Dresselhaus SOC terms, and Zeeman
splitting, respectively. Here
$K_\varphi=r_0^{-1}(-i\partial/\partial\varphi+\Phi/\Phi_0)$ with
$\Phi=\pi r_0^2B$ being the flux of the magnetic field ${\bf B}$,
and $\Phi_0=2\pi\hbar c/|e|$ is the flux quantum, $m$, $\alpha$,
$\beta$, $g$ and $r_0$ are the effective mass, Rashba and $k$-linear
Dresselhaus SOC parameter, effective g-factor and QR radius,
respectively. Here we use the polar Pauli matrices
$\sigma_r=\cos\varphi\sigma_x+\sin\varphi\sigma_y$,
$\sigma_\varphi=-\sin\varphi\sigma_x+\cos\varphi\sigma_y$, that
connected with usual Cartesian Pauli matrices $\sigma_i$
($i=x,y,z$). The constant energy shift that is due to dimension
quantization is omitted in Eq. (\ref{Hamiltonian}).

\begin{figure}
\includegraphics[width=80mm]{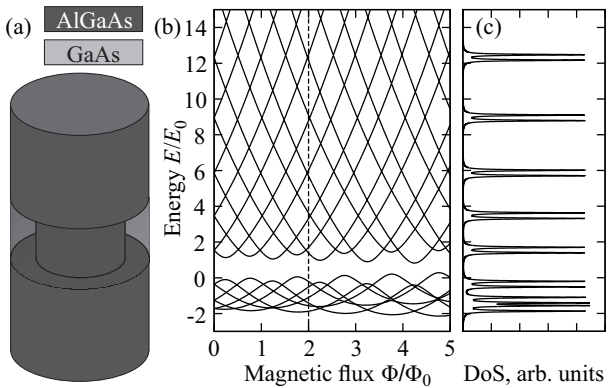}
\caption{\label{fig1} (a) Sketch of one possible realization of
NWQR. (b) Energy spectrum of NWQR as a function of magnetic flux.
$m=0.026m_0$, $2m\alpha r_0/\hbar^2=2.5$, $2m\beta r_0/\hbar^2=0.8$,
$g=-3.0$. (c) Density of states at $\Phi=2\Phi_0$ marked by dashed
line at spectrum, $\gamma=0.03E_0$ with $E_0=\hbar^2/2mr_0^2$.}
\end{figure}

The spectral problem for Hamiltonian (\ref{Hamiltonian}) can be
solved only numerically, but any appropriate set of basis functions
can be used for numerical diagonalization. One can see that the
Hamiltonian without penultimate term commutes with $z$-projection of
total angular momentum,
$j_z=-i\hbar\partial/\partial\varphi+(\hbar/2)\sigma_z$ and
therefore it can be diagonalized analytically. Thus, it is
convenient to use the eigenfunction of such reduced Hamiltonian for
diagonalization of Hamiltonian (\ref{Hamiltonian}). The result of
numerical diagonalization is depicted in Fig.~\ref{fig1}b, where we
restrict ourselves by $40\times 40$ matrix that ensures the perfect
precision for depicted levels. The numerical calculation was
performed with material parameters that are typical for InAs-based
structures for which SOC-effects are more pronounced. One can see
the presence of `gap' (anticrossings) at the spectrum that is due to
penultimate term in Hamiltonian (\ref{Hamiltonian}). It should be
noted, that the energy levels will be $\Phi_0$-periodic at $g=0$.

Additionally, we can calculate the one-particle density of states
(DoS) (see Fig.~\ref{fig1}c) using the Green's function of
Hamiltonian
\begin{equation}
{\rm DoS}(E)=-\frac{1}{\pi}{\rm Im Tr}(E-H+i\gamma)^{-1},
\end{equation}
where $\gamma$ describes the level broadening.

\section{Persistent currents}

Since we know the energy spectrum, then we can use Eq.~(\ref{PC}) to
find PC numerically. However, it is convenient to use the
equilibrium density matrix formalism, and to calculate PC by using
equation
\begin{equation}
\label{current}
I=\frac{e}{2\pi r_0}{\rm Tr}[v_\varphi f_0(H,\mu,T)],
\end{equation}
where $f_0(E,\mu,T)$ is the Fermi distribution function, $E$, $\mu$
and $T$ are the energy, chemical potential and temperature,
respectively. This approach is more general and permits to find
current at finite temperature. In this case we have to know the
matrix elements of velocity operator $v_\varphi$ that in turn can be
found from Heisenberg equation of motion,
$v_\varphi=-(i/\hbar)[r_0\varphi,H]$.

\begin{figure}
\includegraphics[width=80mm]{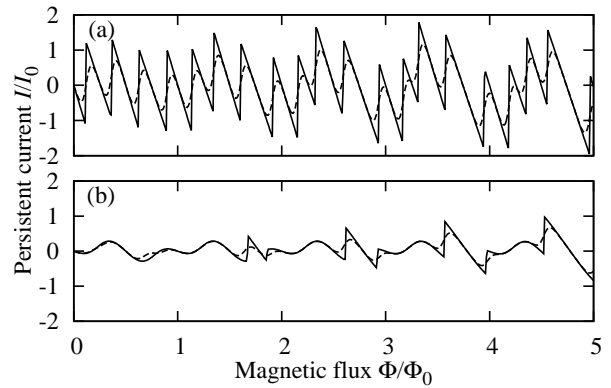}
\caption{\label{fig2} PC as a function of magnetic flux at constant
chemical potential, $m=0.026m_0$, $2m\alpha r_0/\hbar^2=2.5$,
$2m\beta r_0/\hbar^2=0.8$, $g=-3.0$. $I_0=|e|\hbar/4\pi mr_0^2$. (a)
$\mu=4.0E_0$; (b) $\mu=0.0$. Full line corresponds to $T=0$, dashed
line corresponds to $T=0.1E_0$.}
\end{figure}

There are possible two different situations: (i) the constant
particle number, $N={\rm const}$, that e.g. can be realized in QR
defined in core-shell NW-structure and (ii) the constant chemical
potential, $\mu={\rm const}$, that can be realized in
electrostatically confined TEG in InAs-NW. In the first case one has
to additionally find the dependence $\mu(\Phi)$ using well-known
relation, $N={\rm Tr}[f_0(H,\mu,T)]$. In the case of $\mu={\rm
const}$ the abrupt change in PC occurs near the crossing of energy
levels with the chemical potential, whereas in $N={\rm const}$ case
it happens close to level crossing (anticrossing).

The numerical calculation of PC as a function of magnetic flux at
$\mu={\rm const}$ is depicted in Fig.~\ref{fig2}. One can see the
usual saw-toothed behavior for the high chemical potential
(Fig.~\ref{fig2}a) at zero temperature, but there is no
$\Phi_0$-periodicity due to non-zero g-factor. The increasing of the
temperature leads to smoothing of $I(\Phi)$-dependence. When the
chemical potential lies in a `gap' the PC behavior is significantly
different. In this case $I(\Phi)$-dependence is smoothly oscillatory
even at $T=0$ (see Fig.~\ref{fig2}b).

It is interesting to study persistent spin current (PSC) as well.
Usually, the spin current is the pseudo-tensor $I^i_j$
($i,j=x,y,z$), that components describe the spin component $s_i$
carrying in $j$-th spatial direction. In our case of thin QR there
is only one coordinate direction, tangential to QR circumference,
i.e. in our case we deal with pseudo-vector $I^i_\varphi\equiv I^i$
($i=r,\varphi,z$). Here we use the simple definition of spin-current
operator and PSC can be found from Eq.~(\ref{current}) with
replacement

\begin{equation}
\label{spin_current}
ev_\varphi\rightarrow\frac{\hbar}{4}(\sigma_iv_\varphi+v_\varphi\sigma_i).
\end{equation}

Numerical calculations show that at specific values of magnetic
field and chemical potential there is possibility to observe a pure
PSC, i.e. in this case charge current is zero (see for instance Ref.
\cite{Splettstoesser2003}). The realization of pure PSC is
convenient with help of electric gates, that can control not only
electron concentration, but the SOC parameter $\alpha$ and in a less
degree $\beta$.

\section{Conclusion}

In conclusion, we proposed two models of QR defined in zinc-blende
NWs and constructed the one-particle Hamiltonian for electron in
NWQR taking into account both Rashba and $k$-linear Dresselhaus SOC.
The specificity of spin-orbit terms in [111]-oriented QR manifests
in energy levels and PCs. The energy levels as a function of
magnetic field are found revealing several anticrossings that are
due to anisotropic part of Dresselhaus SOC. Using the equilibrium
density matrix formalism PC and PSC are calculated. Charge and spin
PC behavior differs from conventional one, especially for carrier
concentration corresponding to the Fermi level position close to
anticrossings. The variation of magnetic field or carrier
concentration and SOC strength by means of gate can lead to pure
spin PC when the charge current is zero.

\end{document}